\documentclass{mn2e}
\usepackage{graphicx}
\usepackage{amsbsy}
\usepackage{amsmath}
\usepackage{epsf}
\usepackage{paralist}
\usepackage{makecell}
\usepackage{upgreek}

\newcommand{\beq}{\begin{equation}}
\newcommand{\eeq}{\end{equation}}

\DeclareMathAlphabet{\mathsfsl}{OT1}{cmss}{bx}{sl}
\SetMathAlphabet{\mathsfsl}{bold}{OT1}{cmss}{bx}{sl}

\usepackage{amssymb}
\usepackage{pifont}

\usepackage{epstopdf}
\usepackage{color}        

\begin{document}

\title[The 3D shape of Musca]
{The Musca molecular cloud: The perfect ``filament" is still a sheet}

\author[Tritsis et al.]
  {A.~Tritsis$^{1}$\thanks{E-mail:atritsis@uwo.ca}, F.~Bouzelou$^{2, 3}$, R.~Skalidis$^{2, 3}$,  K.~Tassis$^{2, 3}$, T.~En{\ss}lin$^{4, 5}$, G.~Edenhofer$^{4, 5}$ \\
    $^1$Department of Physics and Astronomy, University of Western Ontario, London, ON N6A 3K7, Canada \\
    $^2$Department of Physics and ITCP, University of Crete, Voutes, 70013 Heraklion, Greece \\
    $^3$Institute of Astrophysics, Foundation for Research and Technology-Hellas, Voutes, 70013 Heraklion, Greece \\
    $^4$Max Planck Institute for Astrophysics, Karl-Schwarzschildstra{\ss}e 1, D-85748, Garching, Germany \\
    $^5$Ludwig-Maximilians-Universit\"{a}t, Geschwister-Scholl Platz 1, D-80539, Munich, Germany}
\maketitle

\begin{abstract}

	The true 3-dimensional (3D) morphology of the Musca molecular cloud is a topic that has received significant attention lately. Given that Musca does not exhibit intense star-formation activity, unveiling its shape has the potential of also revealing crucial information regarding the physics that dictates the formation of the first generation of stars within molecular clouds. Here, we revisit the shape of Musca and we present a comprehensive array of evidence pointing towards a shape that is extended along the line-of-sight dimension: (a) 3D maps of differential extinction; (b) new non-local thermodynamic equilibrium radiative transfer simulations of CO rotational transitions from a sheet-like, magnetically-dominated simulated cloud; (c) an effective/critical density analysis of available CO observations; (d) indirect consequences that a filamentary structure would have had, from a theoretical star-formation perspective. We conclude that the full collection of observational evidence strongly suggests that Musca has a sheet-like geometry.  

\end{abstract}



\section{Introduction}\label{intro}

Quiescent molecular clouds such as Musca (Pereyra \& Magalh{\~a}es 2004; Cox et al. 2016; Kainulainen et al. 2016; Hacar et al. 2016) and Polaris (Panopoulou et al. 2016) with little, to no active star formation, hold a special place in our efforts of formulating a predictive theory for the formation of stars. Without any internal sources of energy, observations of such clouds are easier, albeit non-trivial, to interpret, and to therefore identify the dominant processes controlling their formation and evolution.

These processes are also expected to manifest in the 3-dimensional (3D) morphologies of molecular clouds. If the magnetic field that permeates them is dynamically unimportant then filamentary/prolate clouds are formed as a result of turbulence (e.g., Andr{\'e} et al. 2010; Federrath 2016). On the other hand, if magnetic fields play an important role in the support of molecular clouds against their self gravity and are responsible for the low efficiency of star formation (e.g., Mouschovias, Tassis, \& Kunz 2006), clouds are expected to have oblate or sheet-like geometries with the mean magnetic field oriented parallel to the shortest axis of the cloud (e.g., Mouschovias 1978; Tritsis et al. 2015; Skalidis et al. 2021). Consequently, determining the 3D shapes of clouds could provide important clues as to the role of magnetic field in star formation.

The question of the shape of clouds and cores has been approached statistically by many studies over the past three decades (e.g., Curry 2002; Jones et al. 2001; Jones \& Basu 2002; Tassis 2007; Tassis et al. 2009; Lomax et al. 2013). Unfortunately, given that we can only observe the 2-dimensional (2D) projections of clouds, finding their 3D shapes is a challenging and non-trivial task, even via statistical means. As a result, the picture in the literature, as to whether clouds have primarily prolate or oblate shapes, is far from clear. Early efforts of the past decade to estimate the 3D morphology on a cloud-by-cloud basis (Li \& Goldsmith 2012) relied on spectral observations and  radiative-transfer analysis to obtain the number density of the cloud. Together with column-density measurements these number-density estimates can directly yield the line-of-sight (LOS) dimension of the cloud. However, the radiative-transfer and chemical processes that such analyses are based on, involve large uncertainties and can be highly degenerate.

\begin{figure}
\includegraphics[width=1.025\columnwidth, clip]{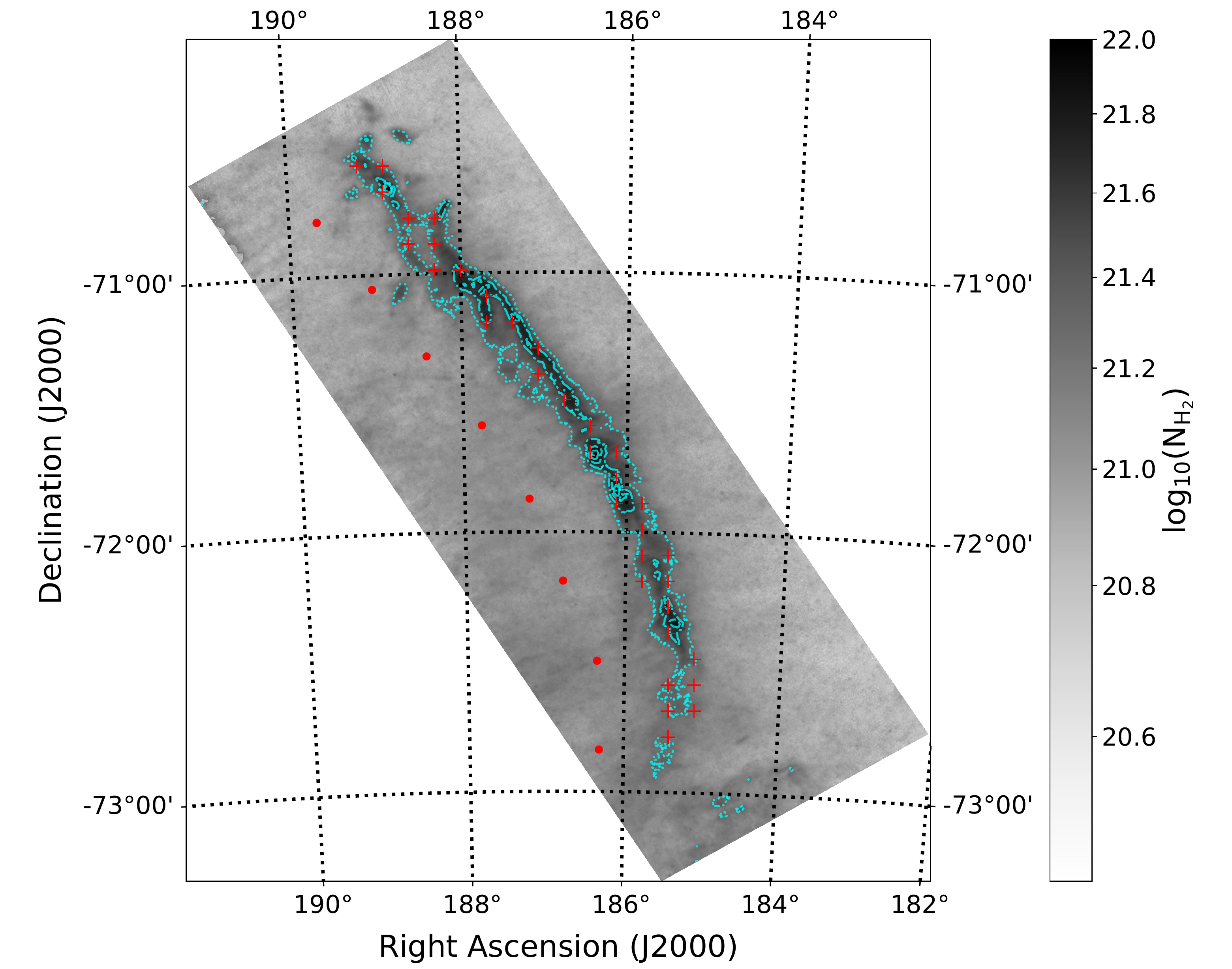}
\caption{Column density map of the Musca molecular cloud from Cox et al. (2016). With the red crosses and points we mark the sightlines we considered to estimate the cloud's depth based on dust extinction maps (see text). The dotted, dash-dotted, dashed and solid light-blue contours mark regions with column densities higher than $2\times 10^{21}$, $4\times 10^{21}$, $6\times 10^{21}$ and $8\times 10^{21}$~$\rm{cm^{-2}}$, respectively.
\label{MuscaImage}}
\end{figure}

The first claim of an accurate determination of the 3D shape of a molecular cloud as a whole came by Tritsis \& Tassis (2018). On the basis that the striations (Tritsis \& Tassis 2016) surrounding the Musca molecular cloud (see Fig.~\ref{MuscaImage}) are formed as a result of trapped hydromagnetic waves, Tritsis \& Tassis (2018) performed a normal-mode analysis and were able to determine the hidden, LOS dimension of the cloud. They found that Musca has a sheet-like geometry instead of a cylindrical one as was the common wisdom at the time (Kainulainen et al. 2016; Hacar et al. 2016; Cox et al. 2016).

This result was recently criticized by Bonne et al. (2020a) and Bonne et al. (2020b) who performed a radiative-transfer analysis of CO isotopologue observations using the publicly-available code $\textsc{RADEX}$ (van der Tak et al. 2007) to estimate the number density of the cloud. Combining their estimation for the density with column-density maps from \textit{Herschel} (Cox et al. 2016) they found that the depth of the cloud ranges from $\sim$0.1 to $\sim$0.7 pc within 0.2 pc from the ``crest" of Musca. Even more recently, Zucker et al. (2021) analysed the 3D dust-extinction maps by Leike et al. (2020) and concluded that Musca has a true filamentary shape in 3D space. However, this statement is highly surprising given that the width (short axis) of the dense structure of Musca (Fig.~\ref{MuscaImage}) on the plane of the sky is $\sim$0.15 pc (Cox et al. 2016) while the LOS resolution of the 3D dust map from Leike et al. (2020) is 1 pc. Furthermore, as Leike et al. (2020) point out, the resolution limit for a 3D reconstruction to be considered reliable is 2 pc. In other words, if the dense structure of Musca was indeed filamentary, then this structure should not be resolved/observed in the 3D dust maps, contrary to what is observed in these data (see the following section).

In this study, we discuss a number of observations towards the Musca molecular cloud and demonstrate that, collectively, the available observations suggest that Musca has significant LOS depth. In \S~\ref{dustmaps} we analyze the 3D differential-extinction maps from Leike et al. (2020). In \S~\ref{rt} we compare non-local thermodynamic equilibrium (non-LTE) radiative transfer simulations of the sheet-like geometry proposed by Tritsis \& Tassis (2018) with available observations and in \S~\ref{coCrit} we present simple effective/critical density arguments to provide lower limits on the LOS of Musca. We discuss our findings in the context of star-formation theories and conclude in \S~\ref{discuss}.

\begin{figure*}
\includegraphics[width=2.0\columnwidth, clip]{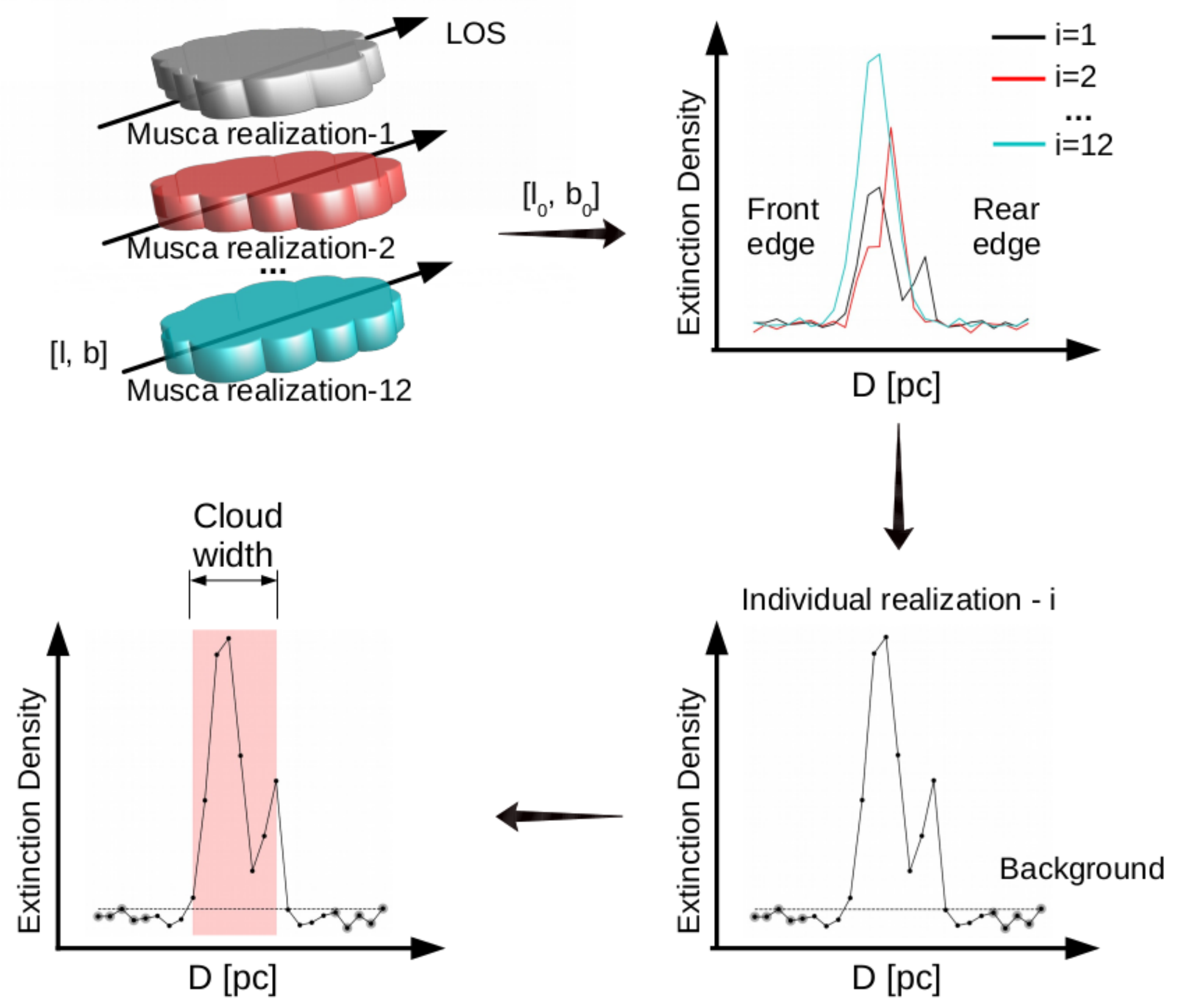}
\caption{Description of our analysis for the shape of Musca based on 3D data of differential dust extinction as a function of distance (Leike \& En{\ss}lin 2019; Leike et al. 2020). We start by selecting a line-of-sight (LOS; upper left). For each LOS selected there are 12 individual realizations of the differential extinction (upper right). For each of these individual realizations we define a background based on the highest 10\% of 5 points in front and 5 points behind the cloud (gray-shaded points in lower right and lower left panels). The LOS width of the cloud is then measured as the interval in which the differential extinction curve is continuously larger than the background (lower left panel). We repeat the same process for each individual realization to create a distribution of widths for each LOS selected.
\label{DustRedAnalysis}}
\end{figure*}


\section{The shape of Musca as seen in 3D maps of differential extinction}\label{dustmaps}

\subsection{Data analysis}\label{methodology}

Recently, Leike \& En{\ss}lin (2019) and Leike et al. (2020) used the catalog compiled by Anders et al.(2019) based on data from Gaia (Gaia Collaboration et al. 2018), 2MASS (Skrutskie et al. 2006), Pan-STARRS (Kaiser et al. 2002) and ALLWISE (Wright et al. 2010) to produce all-sky 3D maps of differential extinction in the G-band within $\sim$400 pc from the Sun with a 1 pc resolution. Here, we use this 3D map which is part of the \textsc{dustmaps} package (Green 2018) in order to study the shape of Musca.

We begin our analysis by selecting 34 sight-lines on the ridge of Musca, marked with the red crosses on Fig.~\ref{MuscaImage}. In order to examine whether the dense structure in Musca is connected with its more diffuse parts where striations are observed (Tritsis \& Tassis 2018) we considered eight additional sightlines below the main ridge of the cloud (red dots on Fig.~\ref{MuscaImage}). All sightlines considered on the dense structure of Musca are equidistant from each other and are separated by $\sim$0.3 pc.

Our methodology is summarized in Fig.~\ref{DustRedAnalysis}. For each sightline selected there are 12 different realizations of the differential dust extinction as a function of distance (upper right panel of Fig.~\ref{DustRedAnalysis}). These different realizations are posterior samples generated by the 3D dust reconstruction algorithm used in Leike et al. (2020). For each of these different realizations we consider five points in front and five points behind the cloud, in the distance range [160, 164] pc and [181, 185] pc, respectively. We use these ten points to define a ``background level" as follows. We find the point with the second largest value (see lower right panel of Fig.~\ref{DustRedAnalysis}) as this represents the 90\% confidence level for the value of the background (in Appendix~\ref{background} we show that our choices for defining the background do not qualitatively or quantitatively affect our results). This constitutes a conservative choice, as opposed, for instance, to considering an average value from these ten points. However, we wish to ensure that no background points are included when we measure the depth of Musca. Another approach for defining the background level would be to consider the mean and 3 times the standard deviation of the ten points we use to define the background. Such an approach would be more suitable if the background points followed a Gaussian distribution which, however, is not the case here. Nevertheless we have repeated our analysis using this approach and present our results in Appendix~\ref{ThreeSigma}. Finally, we measure the depth of the cloud as the interval where the differential extinction is \textit{continuously} higher than the background. Finally, for each LOS, we report the median and 16 and 84 percentiles from all 12 different realizations. 

This analysis has the advantage that our final results on the depth of Musca do not depend on fitting some parametric model (as done by Zucker et al. 2021). In fact, in Appendix~\ref{fitting} we present the outcome of fitting three different parametric models to the dust-extinction data; a simple single Gaussian function, two Gaussian components and a super-Gaussian function; and demonstrate that neither is a good representation of the data.

\subsection{The depth of Musca}\label{widthDustSec}

\begin{figure*}
\includegraphics[width=2.0\columnwidth, clip]{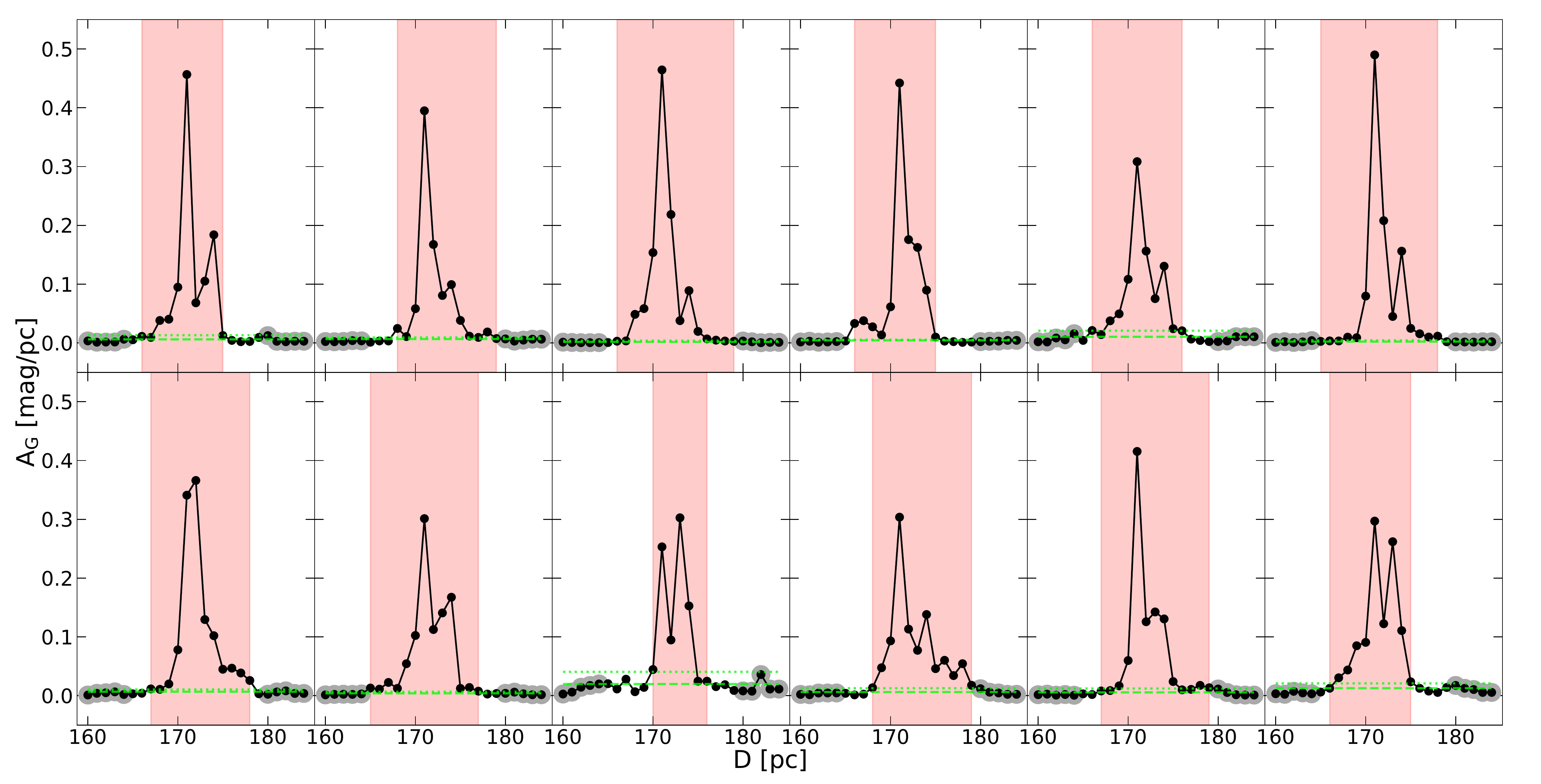}
\caption{Twelve different realizations of the differential extinction as a function of distance towards a random LOS on the ridge of Musca. The gray-shaded points are the points used for defining the value of the background (marked with the green dashed line), and the red-shaded region shows the derived LOS depth of Musca for each of these different realizations. The dotted green line shows the value of the background if we use a 3$\sigma$ approach for defining it (see text for details).
\label{RealiRidge}}
\end{figure*}

\begin{figure*}
\includegraphics[width=2.0\columnwidth, clip]{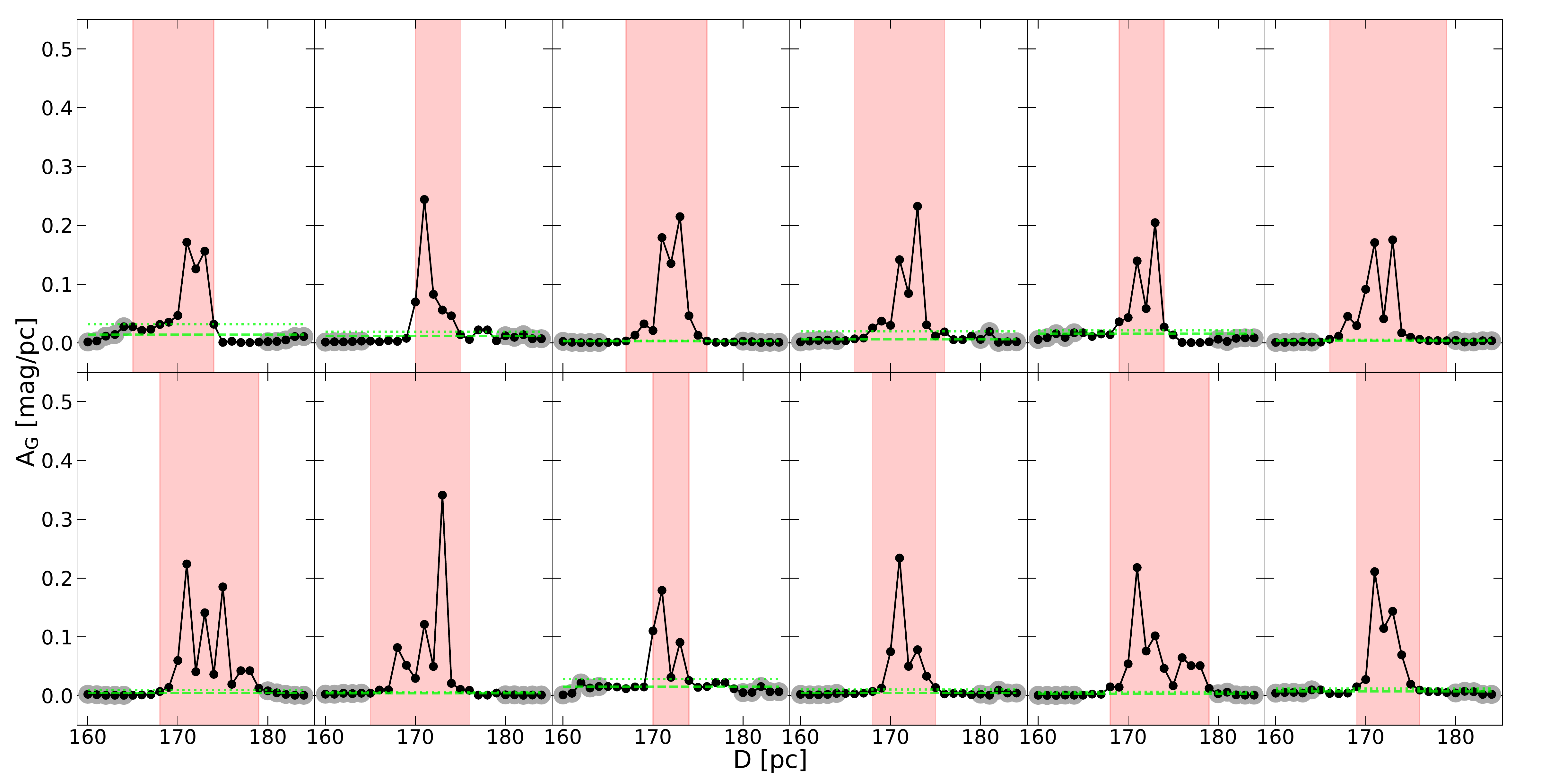}
\caption{Same as in Fig.~\ref{RealiRidge} but for a LOS on the striations region of Musca.
\label{RealiStriations}}
\end{figure*}

In Figs.~\ref{RealiRidge} and ~\ref{RealiStriations} we present examples of 12 different realizations for a sight line on the ridge of Musca and a sightline on the striations region, respectively. As in the two lower panels of Fig.~\ref{DustRedAnalysis}, the points used for defining the background are marked as gray-shaded points. The value of the background is marked with the green dashed line. We show the measured LOS depth of Musca for each realization with the red shaded region. 

In the upper panel of Fig.~\ref{DustDepthfLOS} we present our results for the depth of Musca for each of the sightlines considered on the ridge of the cloud, moving from north to south. Similarly, in the bottom panel of Fig.~\ref{DustDepthfLOS}, we show our results for the depth of Musca in the striations region. In both cases, the depth of the cloud is estimated to be larger than 6 pc, suggesting that Musca is extended along the LOS. These results also suggest that striations and the dense structure are connected.

\subsection{Distance to Musca}\label{distDustSec}

We also estimate the distance at which the cloud is located. To this end, we measure the distance to the front edge of the cloud, i.e. the first point where the dust extinction is continuously higher than the value of the background. In Fig.~\ref{distDust} we show the median value for the distance from the different realizations together with the 16 and 84 percentiles. We find that the cloud is located at a distance of $\sim$~165-170 pc. This result is in excellent agreement with previous measurements for the distance of Musca (Knude 2010).

\begin{figure}
\includegraphics[width=1.0\columnwidth, clip]{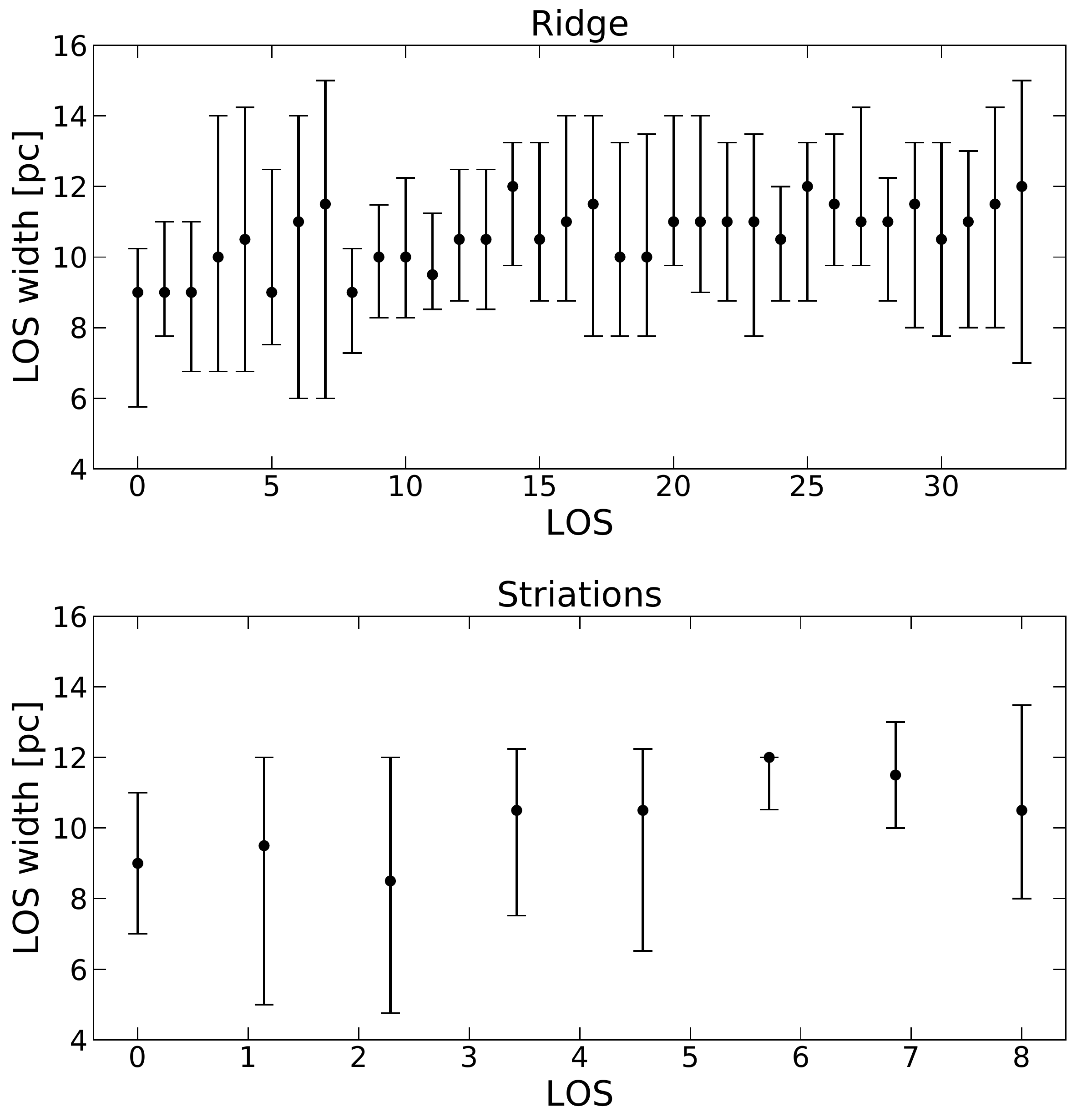}
\caption{The LOS width retrieved on the ridge (upper panel) and on striations (bottom panel) of Musca for the sightlines shown in Fig.~\ref{MuscaImage}. The points and errorbars represent the median and 16 and 84 percentiles from the 12 different realizations for each LOS. The width is consistently above 6 pc indicating that Musca has a sheet-like rather than a filamentary morphology. Moreover, striations and the dense structure in Musca have similar depths.
\label{DustDepthfLOS}}
\end{figure}

\begin{figure}
\includegraphics[width=1.0\columnwidth, clip]{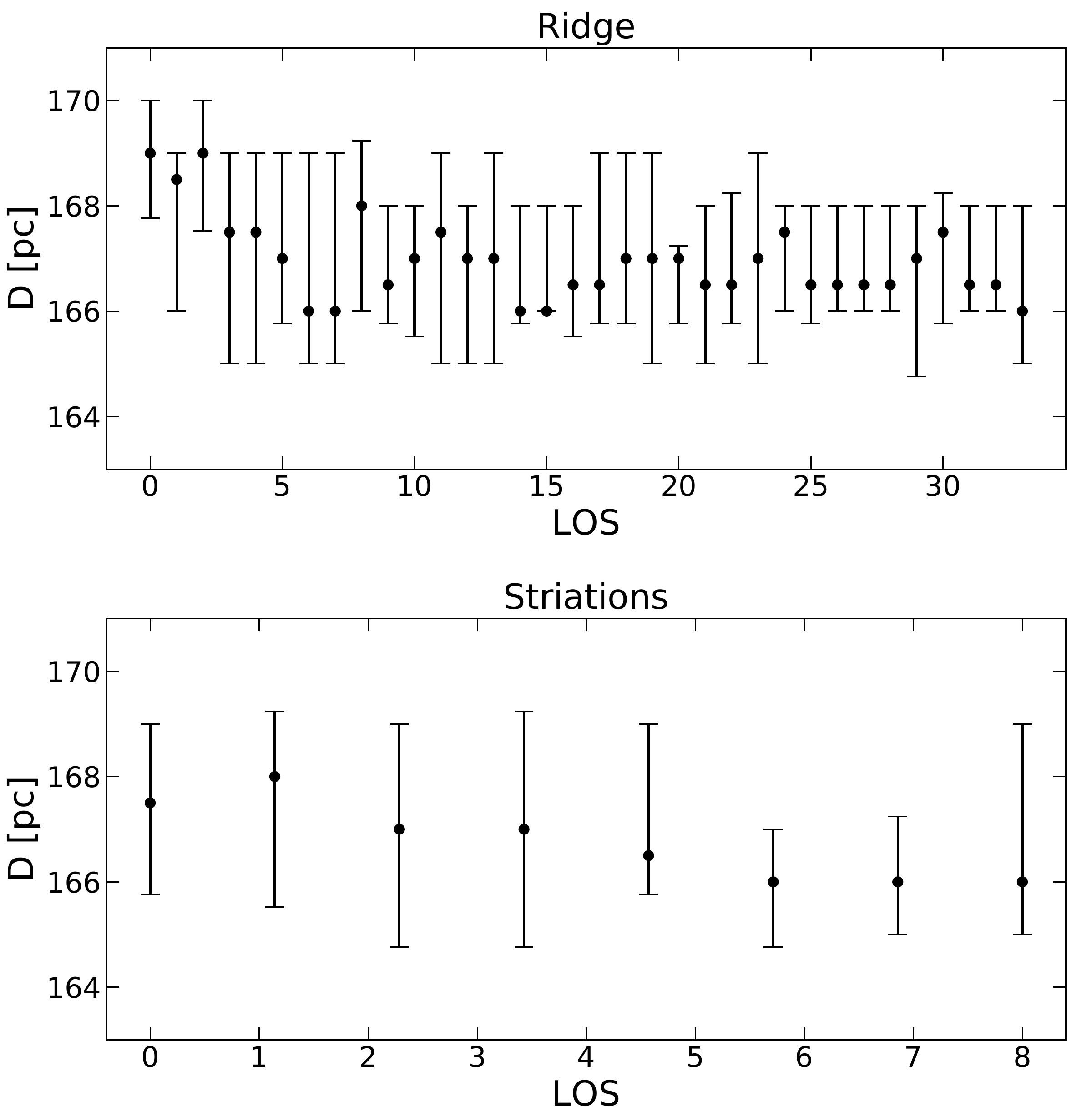}
\caption{Measured distance to the ridge of the Musca cloud (upper panel) and the striation region (lower panel). As in Fig.~\ref{DustDepthfLOS}, the points and errorbars represent the median and 16 and 84 percentiles from the 12 different realizations for each LOS.
\label{distDust}}
\end{figure}

From the results presented in Figs.~\ref{DustDepthfLOS} and~\ref{distDust} we can also safely exclude the possibility that Musca is a filamentary structure that is highly inclined with respect to the POS. That is, if Musca was a cylindrical cloud with a diameter of $\sim$0.1--0.2 pc, then an inclination angle close to 89$^\circ$ would be required for such a diameter to be observed as $\sim$8--10 pc (Fig.~\ref{DustDepthfLOS}). At the same time, such an inclination angle would imply that Musca's length should be more than $\sim$350 pc, for its POS length to be projected as $\sim$10 pc. This would result in significant differences in the distance measured between the two ends of the cloud which, however, is not the case (Fig.~\ref{distDust}).  

\subsection{Caveats and additional considerations}

\begin{figure*}
\includegraphics[width=2.2\columnwidth, clip]{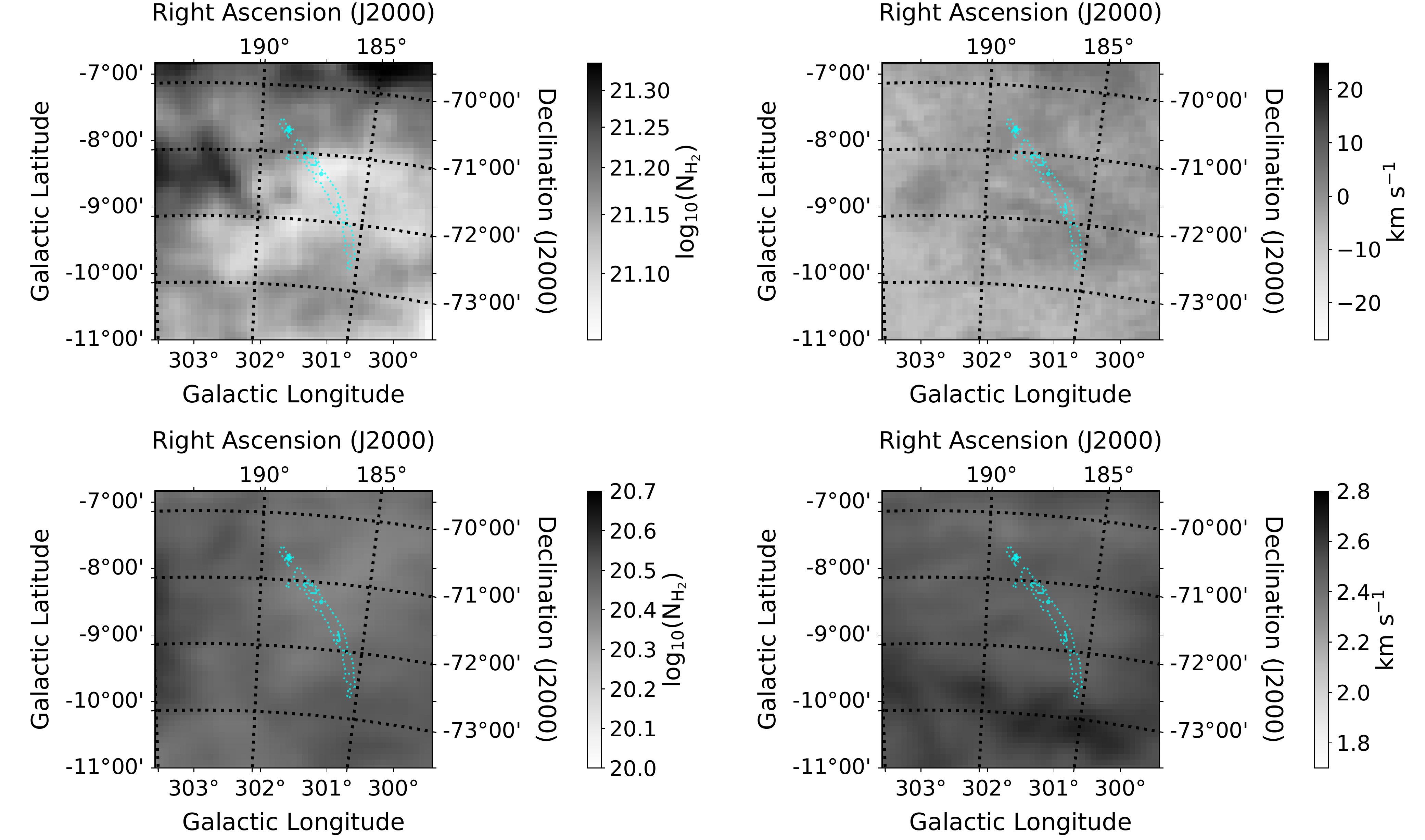}
\caption{In the upper left and right panels show the total HI column density in the region where Musca is located and the corresponding first-moment map, respectively. In the lower left and right panels we show the HI column density and first-moment maps considering a velocity range from 1-4 $\rm{km~s^{-1}}$, i.e. approximately the same velocity interval where Musca is observed (see Fig. 13 from Bonne et al. 2020b). The blue contours are the same as in Fig.~\ref{MuscaImage}.
\label{HIComp}}
\end{figure*}

One reasonable concern regarding the analysis of the 3D dust-extinction maps is that they could potentially be probing diffuse foreground or background H\textsc{I} gas located along the same LOS instead of Musca. In order to examine if this is indeed the case, we plot in Fig.~\ref{HIComp} the column density and first moment maps of H\textsc{I} gas (left and right columns, respectively) using data from \textsc{HI4PI} (HI4PI Collaboration et al. 2016). In the upper row, we show the total H\textsc{I} column density (calculated over the entire velocity range of the \textsc{HI4PI} survey) and the corresponding first moment map at the same region as Musca, whereas in the lower row we only show these two maps for the velocity interval [1, 4] $\rm{km~s^{-1}}$, i.e. the same velocity range where the molecular gas is observed. The location of the main ridge of Musca is marked with the same contours as in Fig.~\ref{MuscaImage}. Firstly, we note that there is no visual correlation between the H\textsc{I} and the molecular gas. More importantly however, there is no increase in the column density of H\textsc{I} (either in the upper or lower row) moving from the striations region to the ridge of Musca that could explain the increase of a factor of $\sim$2 in dust extinction observed in Figs.~\ref{RealiRidge} and~\ref{RealiStriations}. Therefore, the increase in the amplitude of the dust-extinction profiles, can only be explained by a corresponding increase in column density of the molecular gas. Consequently, the depth measured is that of the molecular gas instead of H\textsc{I} gas located along the same sight line and it is therefore unlikely that Musca is a filamentary cloud embedded in significantly extended diffuse H\textsc{I} gas.

\begin{figure*}
\includegraphics[width=2.1\columnwidth, clip]{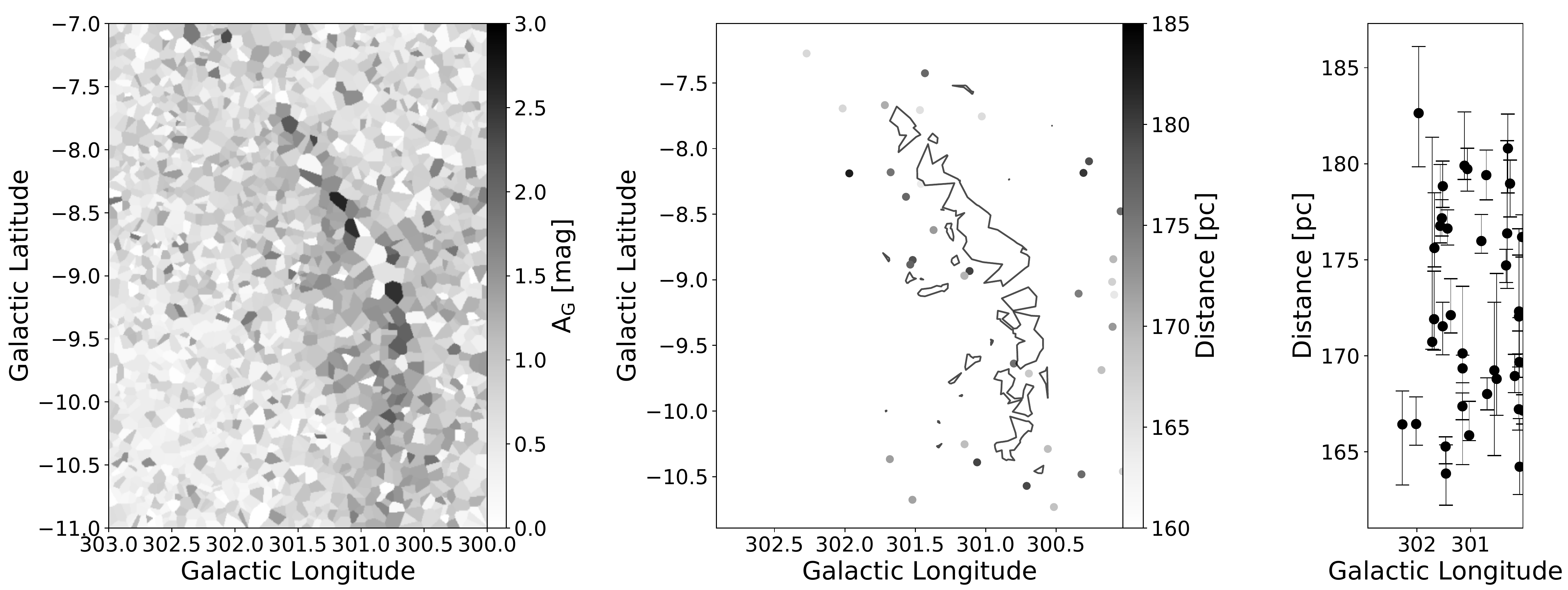}
\caption{Left panel: Voronoi plot of the total G-band extinction from 2187 stars within 500 pc from the Sun. Musca can be clearly identified as a more extinct region. Middle panel: Number of stars within the region Musca is observed and the distance range [165, 180] pc, color coded by their distance. The contour marks the region with $\rm{A_G}\ge1.5$. Right panel: Median distances and uncertainties of stars as a function of Galactic longitude. In the region and distance range of Musca, only 37 stars are available to constrain the cloud's 3D morphology.
\label{GordianFigure}}
\end{figure*}

One possibility for measuring the depth of the cloud in each LOS would be to consider the full width at half maximum (FWHM) for all twelve realizations. However, in a clumpy medium the FWHM would not probe the whole extend of the cloud which is the measurement we are interested in here. Instead, such measurements would only probe a characteristic scale of an overdensity inside the cloud. This point becomes clear by examining Fig.~\ref{RealiRidge}. For approximately 9 out of the 12 realizations, if we consider the FWHM, all the secondary peaks would be completely missed. The resulting measured depth of the total extent of the cloud would then be severely underestimated even though there is significant structure beyond the main peak according to the data. Therefore, clearly the FWHM is not a good proxy for the total extend of the cloud.

Another concern is related to the uncertainties of the 3D dust reconstruction by Leike et al. (2020). Firstly, the positions of the stars have high uncertainties along the LOS (see right panel of Fig.~\ref{GordianFigure}). Additionally, given that these uncertainties are incorporated in the reconstruction algorithm, structures can be smeared out along the LOS. Finally, the uncertainties derived from the posterior samples of the reconstruction by Leike et al. (2020) are only indicative of the real uncertainties, as the used variational inference methodology typically underestimates uncertainties. The situation is further worsened by the fact that only 37 stars are available to provide tight constraints on Musca's 3D shape (see middle and right panels of Fig.~\ref{GordianFigure}). Thus, \textit{claiming that we can exclude the possibility of Musca being a filament on the basis of the 3D dust reconstruction alone, is probably premature.}

That said, for sightlines close to stars where the inner structure of the cloud should be better constrained, multiple ($\gtrsim 60\%$) realizations of the differential extinction exhibit peaks that are more than 4 pc apart. Such peaks in the differential extinction should be data driven. In fact, 55\% of the total realizations for all 34 sightlines on the ridge of Musca exhibit two or more significant peaks in the range where the cloud is located (i.e. inside the red-shaded region shown in Fig.~\ref{RealiRidge}). Here, we define a peak to be significant if its amplitude is at least 10\% that of the maximum amplitude in every given profile. For the striations region, 76\% of the total realizations for all 8 sightlines exhibit 2 or more significant peaks. We emphasize here that the fact that the rest of the profiles do not exhibit multiple peaks does not necessarily imply that Musca has a narrow LOS depth. That is because multiple high-density structures along the LOS (corresponding to peaks in the dust exctinction) may be unresolved due to the resolution of the data and/or there may only be one high-density region embedded in a quasiuniform region with lower density. Therefore, even though the dust-extinction maps do not have significant discriminatory power in regards to whether Musca is a filament or a sheet, the fact that they also seem to suggest a significant depth, is in line with both the normal-mode analysis by Tritsis \& Tassis (2018) and the rest of the observational evidence we provide below.


\section{Radiative transfer simulations}\label{rt}

In this section, we present non-LTE line radiative transfer simulations of $^{12}$C$^{16}$O J = 2$\rightarrow$1, J = 3$\rightarrow$2 and J = 4$\rightarrow$3 transitions using the multilevel line radiative transfer code \textsc{PyRaTE} (Tritsis et al. 2018). We use \textsc{PyRaTE} to post-process the ideal magnetohydrodynamic (MHD) simulation of the sheet-like cloud presented in Tritsis \& Tassis (2018). Here, we re-iterate the initial conditions and main properties of this dynamical simulation but we refer the reader to Tritsis \& Tassis (2018) for more details.

The dimensions of the simulation box were $\rm{L_x} = 10$ pc, $\rm{L_y} = 8$ pc and $\rm{L_z} = 3.5$ pc while the final $x$ and $y$ dimensions of the cloud were 7.7 pc and 5.6 pc, respectively. The initial magnetic field in this simulation is along the $z$ axis. The simulation was carried out on $256^3$ grid. Initially, the number density, and magnetic field strength in the simulation were 100 $\rm{cm^{-3}}$ and 7 $\upmu G$. The evolutionary time when we post-process our simulation with \textsc{PyRaTE} is 2.7 Myrs, i.e. the same output as the one shown in Tritsis \& Tassis (2018). At this stage, the column density of the dense structure formed was found in Tritsis \& Tassis (2018) to be comparable to observations.

The input physical quantities to the \textsc{PyRaTE} code are the density, temperature, velocity field, and molecular number density. However, the simulation outlined above did not contain any non-equilibrium chemical modeling. We therefore simply assume a constant CO number density of $\rm{n_{CO}} = 10^{-3}$ $\rm{cm^{-3}}$ everywhere in the cloud. This value is consistent with the abundance of $\rm{CO}$ calculated in chemodynamical models of (sub)critical magnetized clouds (Tassis et al. 2012). Furthermore, in order to save computational time, the radiative transfer simulation presented here was not carried out using the native resolution presented in (Tritsis \& Tassis 2018) but based on a resampled grid with size $128^3$. The spectral resolution used in this radiative-transfer simulation is 0.05 km~$\rm{s^{-1}}$. Finally, we assume that the simulated cloud is seen edge-on and that the LOS is aligned with the $y$ axis of the simulation.

\begin{figure*}
\includegraphics[width=2.0\columnwidth, clip]{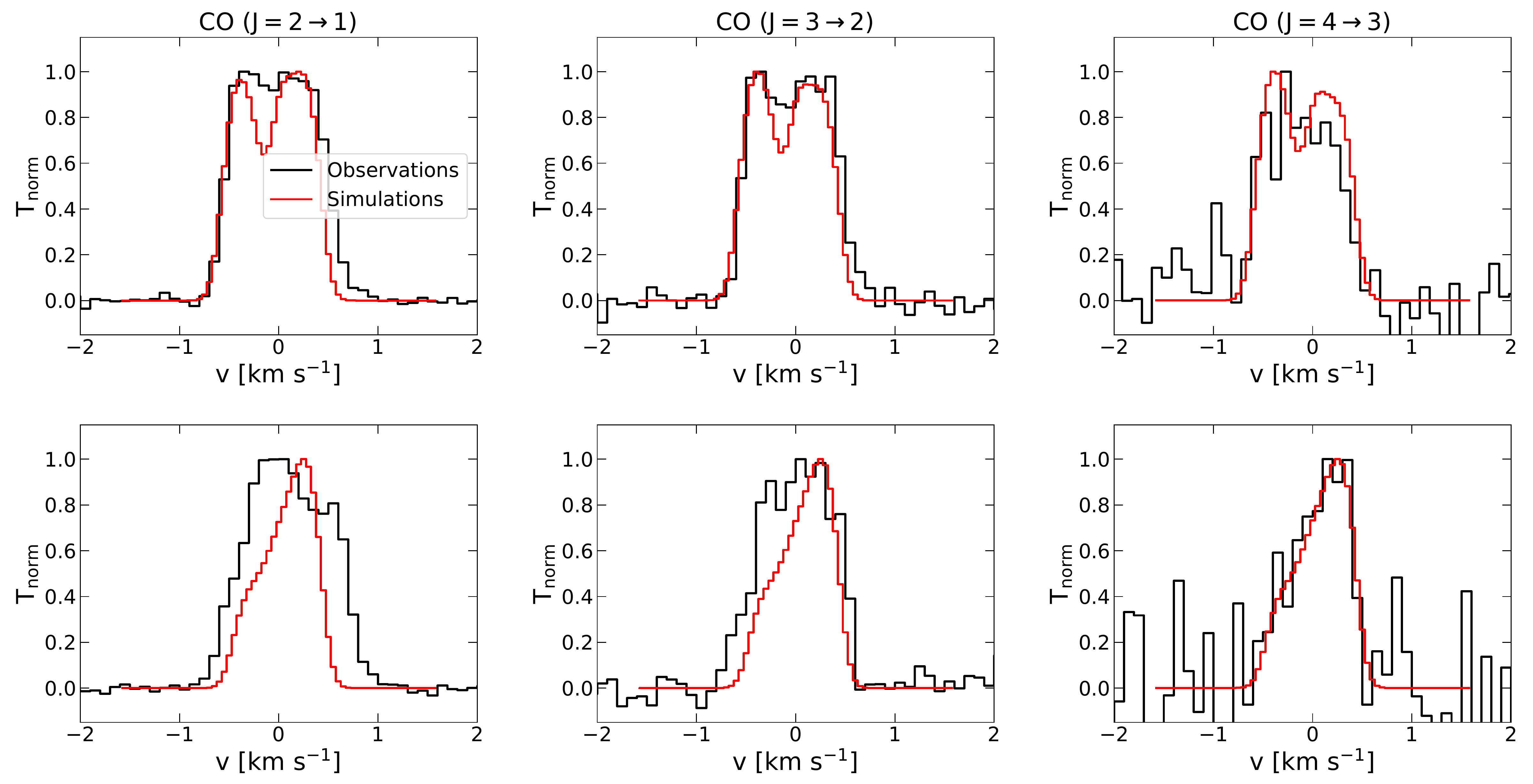}
\caption{Comparison between observed (black lines) and mock-observed (red lines) spectra when a sheet-like simulated morphology is adopted for Musca. In the upper and lower rows we show observed spectra from the north and southern maps, respectively, as defined in Bonne et al. (2020a). From left to right, columns show the $\rm{^{12}CO}$ (J = 2$\rightarrow$1), (J = 3$\rightarrow$2) and (J = 4$\rightarrow$3) transitions of $\rm{^{12}CO}$.
\label{spectra}}
\end{figure*}

In Fig.~\ref{spectra} we compare our simulated spectra with $\rm{^{12}CO}$ observations from Bonne et al. (2020a). The simulated spectra are convolved to the same spectral resolution of 0.1 km~$\rm{s^{-1}}$ as in observations. From the left to the right column we show $\rm{^{12}CO}$ (J = 2$\rightarrow$1), (J = 3$\rightarrow$2), and (J = 4$\rightarrow$3) transitions, respectively. Spectra from the norther and southern maps are shown in the upper and lower rows, respectively (see Bonne et al. 2020a for a definition of these two regions). In order to provide a one-to-one comparison, for each of these two regions, \textit{the different transitions shown are from the exact same physical location}. Likewise, the simulated spectra shown are extracted from the same pixel. Furthermore, the pixels used from the simulation where equidistant, lengthwise, from the edges of the dense structure (approximately as in observations). Both in simulations and observations the spectra shown in the upper rows were at a distance of $\sim$~0.2 pc from the ridge of the dense structure (using our updated distance measurement from \S~\ref{dustmaps}) while for the lower row we considered spectra on the ridge of the dense structure.

Fig~\ref{spectra} shows good qualitative and quantitative agreement between the simulated spectra from the sheet-like-cloud model of Musca and observations both in terms of the linewidth and the features seen in the observed spectra, apart from the bottom left panel. This agreement is remarkable given the fact that the dynamical simulation used did not include any actual chemical modeling and was missing important physical effects such as ambipolar diffusion. 

\begin{figure*}
\includegraphics[width=2.0\columnwidth, clip]{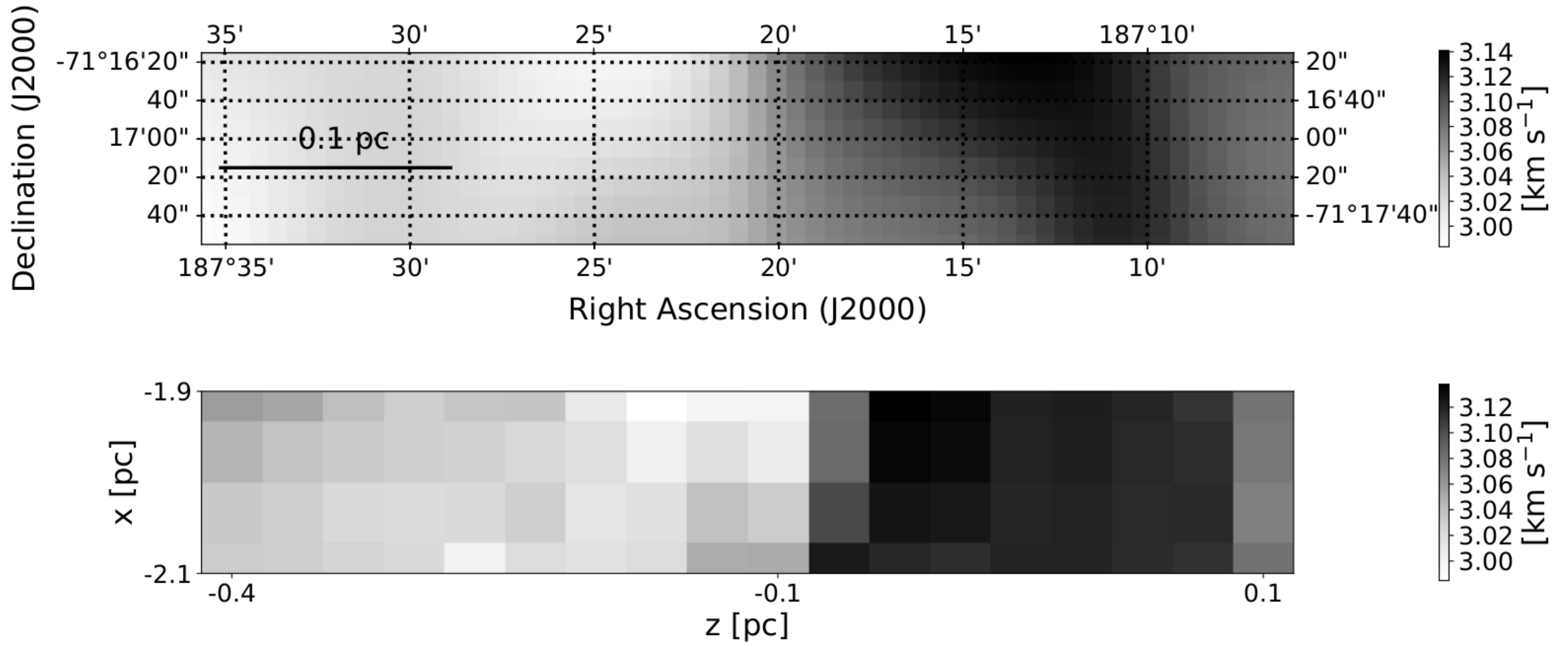}
\caption{Comparison between the observed (upper panel) and simulated (lower panel) first moment maps derived from the $\rm{^{12}CO}$ (J = 2$\rightarrow$1) transition.
\label{fmm}}
\end{figure*}

In the upper and lower panels of Fig.~\ref{fmm} we show, respectively, the observed and our simulated $\rm{^{12}CO}$ (J = 2$\rightarrow$1) first moment map. These two maps extend perpendicularly to the long axis of the denser structure. To facilitate comparison, the two maps are convolved to the same spatial resolution and show the variation in velocity in regions with approximately the same physical size. The simulated map is shifted to the same $\rm{v_{lsr}}$ as in observations.

Our simulated first moment map can reproduce several features seen in the observations. Firstly, the velocity range in both simulations and observations is very similar. More interestingly however, the simulated map can reproduce the fact that the velocity is not monotonically changing when moving across the dense structure. For instance, at z=-0.1 pc (see lower panel of Fig.~\ref{fmm}) the value of the velocity is minimum. On the ridge of the dense structure (z=0 pc) the velocity is increasing but as we then move to z = 0.1 pc the velocity is again decreasing.

Despite the success of our radiative transfer simulations in explaining/reproducing observations, we should note that this agreement between simulated and observed line spectra may not be unique to our theoretical model of a sheet-like cloud. Other models (see Clarke et al. 2018; Bonne et al. 2020a) where the simulated cloud has a filamentary morphology can explain the observed spectra adequately well, albeit these simulations lead to more extended wings in the $\rm{CO}$ spectra in comparison to observations of Musca. Given these results, the question that needs to be asked is: If we cannot break the degeneracy using a forward approach (from simulations to mock-observed spectra), how well can we hope to break the degeneracy using a backwards approach (from spectral observations to the true shape of Musca)?


\section{Inferring Musca's shape from critical \& effective densities of CO isotopologues}\label{coCrit}

A much simpler argument than the radiative-transfer analysis performed by Bonne et al. (2020b) for the density of the cloud can be made using the critical/effective density for excitation of CO isotopologues. The critical density $n_c$ is defined as the density at which a molecular transition is excited assuming no radiative trapping (i.e. no induced excitation or de-excitation). More specifically:
\begin{equation}\label{ncrit}
n_c = \frac{A_{jk}}{\sum_{i<j}C_{ji} + \sum_{i>j}\frac{g_i}{g_j}C_{ij}e^{-(E_i-E_j)/k_BT}}
\end{equation}
(see Shirley 2015 for an extended discussion) where $A_{jk}$ is the Einstein coefficient for spontaneous emission from level $j$ to $k$, $C_{ji}$ and $C_{ij}$ are respectively the de-excitation and excitation coefficients due to collisions\footnote{For molecular-cloud conditions all other collisional partners apart from $\rm{H_2}$ can be safely ignored given that $\rm{H_2}$ is always orders of magnitude more abundant that other chemical species.}, and $g_i$ and $E_i$, and $g_j$ and $E_j$ are the statistical weights and energies of levels $i$ and $j$, respectively. Finally, in Eq.~\ref{ncrit}, $k_B$ is the Boltzmann constant and T is the temperature of the collisional partner.

In Eq.~\ref{ncrit}, the Einstein $B$ coefficient for induced excitation and de-excitation is ignored. However, such an assumption is very rarely valid and is even less appropriate for $\rm{CO}$ which is optically thick. Thus, a much more useful concept is that of the effective density of the cloud (Evans 1999). The effective density is defined as the density required to ``produce" an integrated line intensity of 1 K~km/s, given ``reasonable" assumptions regarding the column density of the molecule in question. The effective density of $\rm{CO}$ (J = 1$\rightarrow$ 0) is 100~$\rm{cm^{-3}}$ (Shirley 2015).

The relation between the concepts of the critical/effective density with the shape of Musca is that, provided a molecular transition is not observed, we can place an upper limit for the density of the cloud. In combination with the column density of the cloud, this upper limit for the number density can yield a \textit{lower limit} for the LOS dimension of the cloud. In Table~\ref{critical} we have calculated the critical density for excitation of CO isotopologues using Eq.~\ref{ncrit} at two different temperatures, 15 and 10 K.

\begin{figure}
\includegraphics[width=1.\columnwidth, clip]{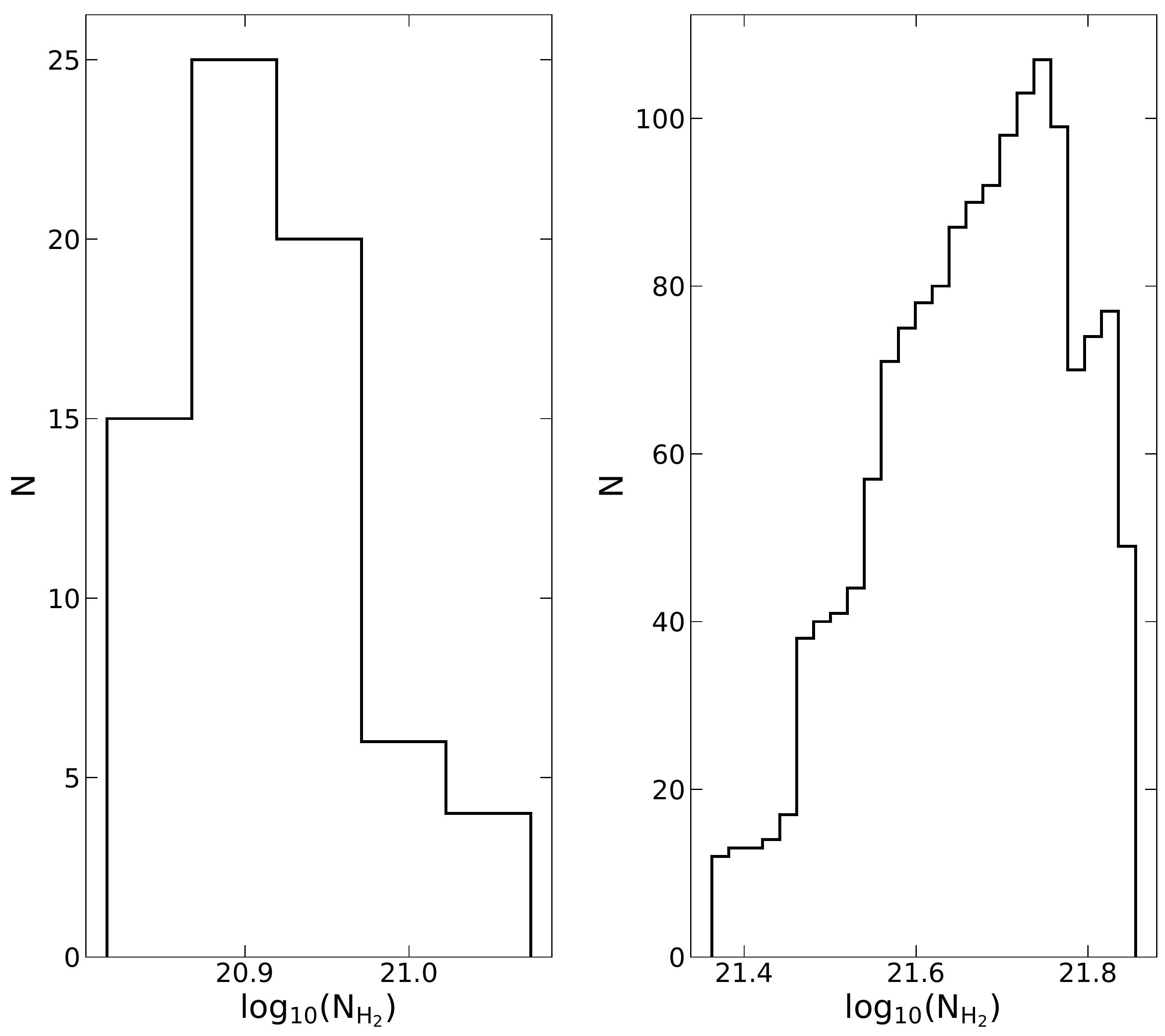}
\caption{Column-density distribution in the striations region (left panel) where the $\rm{^{12}CO}$ (J = 1$\rightarrow$0) and $\rm{^{13}CO}$ (J = 1$\rightarrow$0) transitions are not observed (Mizuno et al. 1998; Mizuno et al. 2001 private communication) and the ridge of the cloud (right) where the $\rm{C^{18}O}$ (J = 1$\rightarrow$0) is not observed (Machaieie et al. 2017). When calculating the distribution shown in the left panel, the \textit{Herschel} column density map was convolved to the same spatial resolution as the radio observations from Mizuno et al. (2001).
\label{cdDistro}}
\end{figure}

Interestingly, in Musca, the $\rm{^{12}CO}$ (J = 1$\rightarrow$0) transition is mostly observed close to the ridge of the cloud. In contrast, there is no significant emission from the striations region (see Fig. 1 from Mizuno et al. 2001 and Fig. 13 from Bonne et al. 2020b). This trend also holds true for the $\rm{^{13}CO}$ (J = 1$\rightarrow$0) emission which is further confined towards the ridge of Musca (see Fig. 1 from Mizuno et al. 1998). In the left panel of Fig.~\ref{cdDistro}, we show the column density distribution in the region where these transitions are not observed, adopting a conservative detection limit of 2$\sigma$. Considering the median value from this distribution	and the effective density of $\rm{^{12}CO}$, we calculate a \textit{lower limit} for the depth of the striations of 2.6 pc. 

Using chemodynamical simulations with a simplified chemical network and radiative transfer post-processing, Seifried et al.(2020) recently found that $\rm{CO}$ bright gas is primarily found at number densities $\rm{n_{H_2}}\gtrsim300~\rm{cm^{-3}}$. Using this estimate for the effective density of the $\rm{^{12}CO}$ (J = 1$\rightarrow$0) transition we find a lower limit for the depth of the striations region of $\sim$0.9 pc. Given that this value is \textit{not} an estimate for the depth of the striations but is rather a \textit{lower limit}, we argue that this result is still inconsistent with a filamentary morphology. Furthermore, given additional constraints for the depth of the dense region in Musca imposed by $\rm{C^{18}O}$ observations (see below), adopting an effective density of $\rm{n_{eff}} = 300~\rm{cm^{-3}}$ for the $\rm{^{12}CO}$ (J = 1$\rightarrow$0) transition, could lead to the counter-intuitive possibility that the depth of cloud is increasing as it gets denser.

Similarly to $\rm{^{12}CO}$ (J = 1$\rightarrow$0) observations, the $\rm{C^{18}O}$ (J = 1$\rightarrow$0) and $\rm{C^{18}O}$ (J = 2$\rightarrow$1) transitions are not observed even on certain regions on the ridge of the cloud (Hacar et al. 2016; Machaieie et al. 2017). The column-density distribution in the region where $\rm{C^{18}O}$ (J = 1$\rightarrow$0) is not observed is shown on the right panel of Fig.~\ref{cdDistro}. If we again consider the median value from this distribution and the critical density of $\rm{C^{18}O}$ (J = 1$\rightarrow$0) from Table~\ref{critical} for a temperature of 10~K, we find a \textit{lower limit} for the depth of the ridge of Musca of 2.1 pc. Thus, we again find that Musca must extend along the LOS over ten times its width on the plane of the sky ($\sim 0.1$ pc) and as a result its 3D shape cannot be that of a filament.

We note here that if we consider the effective density of $\rm{C^{18}O}$ (J = 1$\rightarrow$0) instead of the critical density this lower limit for the depth of Musca will increase. The same holds true if we adopt a slightly higher temperature of $\sim$13 K, as suggested by \textit{Herschel} observations, or relax the unphysical assumption that the density is uniform along the LOS. That is, dividing the total LOS column density assuming a uniform density along the LOS will bias the lower limit for the depth of Musca towards lower values. Conversely, for sightlines where $\rm{CO}$ isotopologue spectra are observed, it is more physical to expect that the emission originates from confined denser condensations along the LOS instead from a uniform-density slab.

\section{Discussion and conclusions}\label{discuss}

We have provided a series of observational arguments on why the true shape of the Musca molecular cloud is that of a sheet-like structure, as was found in Tritsis \& Tassis (2018).

First, we analyzed 3D maps of differential dust extinction by Leike \& En{\ss}lin (2019) and Leike et al. (2020) for several sightlines, both on the dense structure of Musca and the striations region. Based on our analysis, we found that Musca has a depth in the range of 6-15 pc. This value is in good agreement with the value revealed for the depth of Musca based on the normal-mode analysis by Tritsis \& Tassis (2018). We note here that, adopting our new measurement for the distance of Musca (see Fig.~\ref{distDust}), the value found in Tritsis \& Tassis (2018) for the depth of Musca translates to $\sim$7 pc.

We additionally showed that the observed $^{12}$CO transitions towards Musca can be reproduced with non-LTE radiative transfer simulations where the underlying dynamical model is that of a sheet-like structure. We found excellent agreement between the observed and simulated first moment maps derived from $^{12}$CO (J=2$\rightarrow$1), and we demonstrated that a simple critical/effective density analysis of low rotational CO isotopologue transitions strongly suggests a lower limit for the depth of the striations region and the ridge of Musca of at least 2 pc, in agreement with all previous observations.

Collectively, all of these observational pieces of evidence strongly suggest a sheet-like morphology for Musca. From the theoretical perspective, all of these observational considerations can be explained if the cloud is magnetically supported against its self gravity. Molecular spectral observations have demonstrated that Musca is a trans-sonic cloud (Kainulainen et al. 2016; Hacar et al. 2016; Bonne et al. 2020a). Such trans-sonic motions could, at most, support a cloud with mass $\le$100~\(M_\odot\). In contrast, the total mass of Musca is estimated to be $\ge$300~\(M_\odot\) (Mizuno et al. 2001; Cox et al. 2016). Therefore, Musca is most likely \textit{not} supported against its own self-gravity by turbulent motions.

At the same time, we know the cloud is \textit{not} in a state of a global free-fall collapse since the line profiles of all CO isotopologues do not exhibit the infall asymmetries (Evans 1999) characteristic of a collapsing cloud (see Fig. 2 from Hacar et al. 2016; Fig. 6 \& 7 from Bonne et al. 2020a). Consequently, the remaining explanation is that the cloud is not actively forming stars because it is magnetically supported (i.e. it is magnetically subcritical). In turn, if this is indeed the case, then a fragmented sheet-like morphology would be highly favored compared to a filamentary one.

While Musca is one of the first ``filaments" found to actually have a sheet-like morphology, it is not unique. Analysis of Gaia data has shown that other clouds which were thought to have coherent filamentary morphologies, including the Taurus molecular cloud, are in fact extended along the LOS (Gro{\ss}schedl et al. 2018; Fleming et al. 2019).

\begin{table}
\begin{center}
\begin{tabular}{ c c c c}
\hline\hline
 & T = 15 [K] & T = 10 [K] \\
\hline
 Transition & $\text{n}^{thin, no~bg}_{crit}$ & $\text{n}^{thin, no~bg}_{crit}$ \\
 &  [cm$^{-3}$] &  [cm$^{-3}$] \\
\hline 
 $^{12}$C$^{16}$O (J = 1 $\rightarrow$ 0) & 6.7$\times$$10^2$ & 9$\times$$10^2$ & \\  
 $^{13}$C$^{16}$O (J = 1 $\rightarrow$ 0) & 5.7$\times$$10^2$ & 7.6$\times$$10^2$ \\
 $^{12}$C$^{18}$O (J = 1 $\rightarrow$ 0) & 5.6$\times$$10^2$ & 7.5$\times$$10^2$ \\
 $^{12}$C$^{16}$O (J = 2 $\rightarrow$ 1) & 4.8$\times$$10^3$ & 5.5$\times$$10^3$ \\
 $^{13}$C$^{16}$O (J = 2 $\rightarrow$ 1) & 4.2$\times$$10^3$ & 4.7$\times$$10^3$ \\
 $^{12}$C$^{18}$O (J = 2 $\rightarrow$ 1) & 4.1$\times$$10^3$ & 4.7$\times$$10^3$ \\
\hline\hline
\end{tabular}
\end{center}
\caption{\label{critical} In the two first columns we calculated the critical densities for various CO isotopologue transitions for two different temperatures, T = 15 K and T = 10 K. The critical density is calculated as in Shirley (2015).}
\end{table}

\section*{Acknowledgements}

We thank the anonymous referee for comments and suggestions that helped improve this manuscript. We thank Y. Fukui and K. Tachihara for providing the $^{12}$C$^{16}$O (J = 1 $\rightarrow$ 0) observations from Mizuno et al. (2001). We also thank S. Basu, G.V. Panopoulou \& V. Pavlidou for useful comments and discussions. This work was supported by the Natural Sciences and Engineering Research Council of Canada (NSERC), [funding reference \#CITA 490888-16]. KT and RS acknowledge support from the European Research Council (ERC)under the European Unions Horizon 2020 research and innovation programmeunder  grant  agreement  No.  771282. We acknowledge use of the following software: \textsc{Astropy} (Astropy Collaboration et al. 2013; Astropy Collaboration et al. 2018), \textsc{Matplotlib} (Hunter 2007) and \textsc{Numpy} (Harris et al. 2020). Usage of the Metropolis HPC Facility at the CCQCN of the University of Crete, supported  by  the European Union Seventh Framework Programme (FP7-REGPOT-2012-2013-1) under grant agreement no. 316165, is also acknowledged.

\section*{DATA AVAILABILITY}

The data from the radiative-transfer simulations presented herewith are available from the corresponding author upon reasonable request.

\appendix
\section{Derived LOS width considering different backgrounds}\label{background}

\begin{figure*}
\includegraphics[width=2.0\columnwidth, clip]{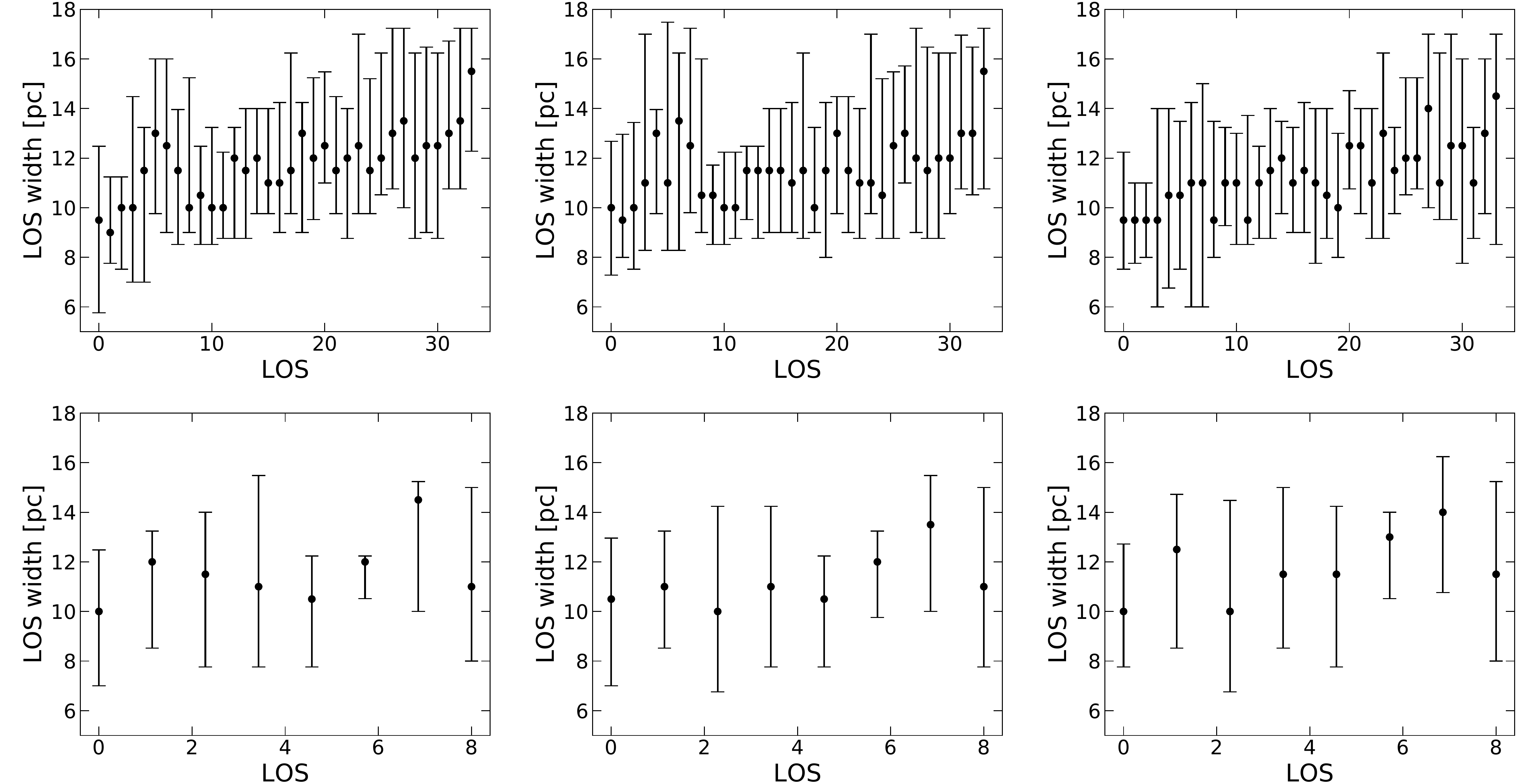}
\caption{Same as in Fig.~\ref{DustDepthfLOS} but considering a different distance range for defining the background (see text). In the upper row we show our results for sightlines on the ridge on Musca while in the bottom row we present our results for the striations region.
\label{DiffBackRes}}
\end{figure*}

Here, we examine whether our choice of the distance range when defining a background affects our results for the derived LOS width of Musca. In the upper and lower rows of Fig.~\ref{DiffBackRes} we present the derived depth for different sightlines on the ridge and the striations region of Musca when considering a distance range [158, 162]$\bigcup$[183, 187] pc (left column), [159, 163]$\bigcup$[184-187] pc (middle column) and [156, 160]$\bigcup$[181-185] pc (right column).

Fig.~\ref{DiffBackRes} demonstrates that while for each individual LOS there is some fluctuation on the median value derived for the depth when selecting a different distance range for defining the background, all points agree within errorbars. Furthermore, regardless of the choice of the distance range for defining the background, no LOS is found to have a depth below $\sim$6 pc. This result is again consistent with Musca being a sheet-like cloud seen edge on.

\section{Alternative background level definition}\label{ThreeSigma}

\begin{figure}
\includegraphics[width=1.0\columnwidth, clip]{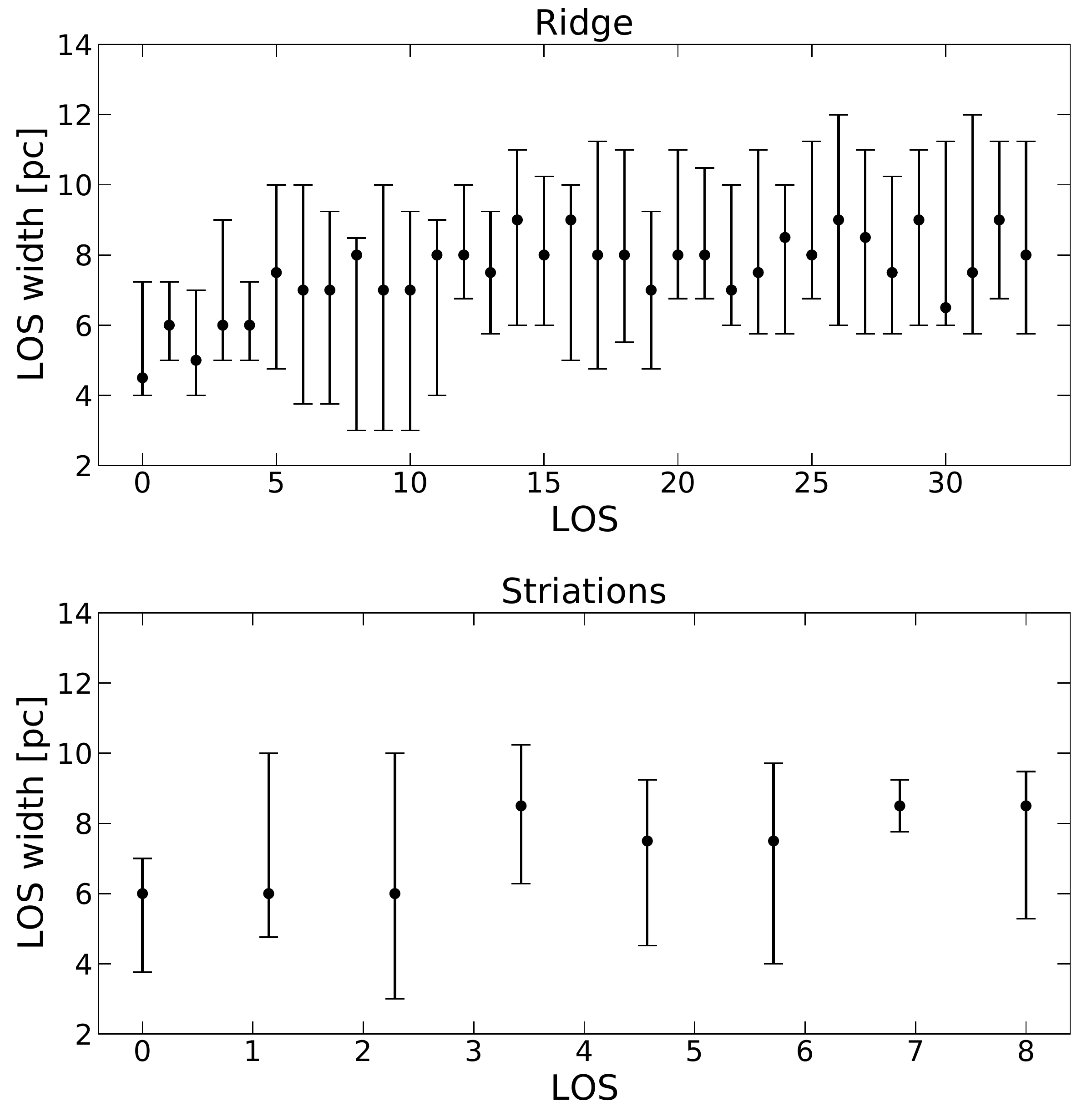}
\caption{Same as in Fig.~\ref{DustDepthfLOS} but considering a different definition for calculating the background level. In this definition, the background level is calculated based on the mean and 3 times the standard deviation of the ten points considered from the near and far sides of the cloud.
\label{LOSwidth_fLOS_3sigmaApproach}}
\end{figure}

In Fig.~\ref{LOSwidth_fLOS_3sigmaApproach} we present our results when we consider an alternative approach for calculating the background level. In this approach, we consider the mean and three times the standard deviation of the ten points in the distance range [160, 164]$\bigcup$[181, 185] pc. As already mentioned in \S~\ref{methodology}, this approach can potentially be problematic when the background data points do not follow a Gaussian distribution, as it can lead to the background level being higher than the maximum value of all ten points (see for instance the third panel from the left in the bottom row of Fig.~\ref{RealiRidge}). Nevertheless, adopting this approach for the definition of the background does qualitatively change our estimate for the depth of Musca, since the depth of the cloud is never found to be less than 3 pc.

\section{Fitting a parametric model to differential extinction}\label{fitting}

\begin{figure*}
\includegraphics[width=2.0\columnwidth, clip]{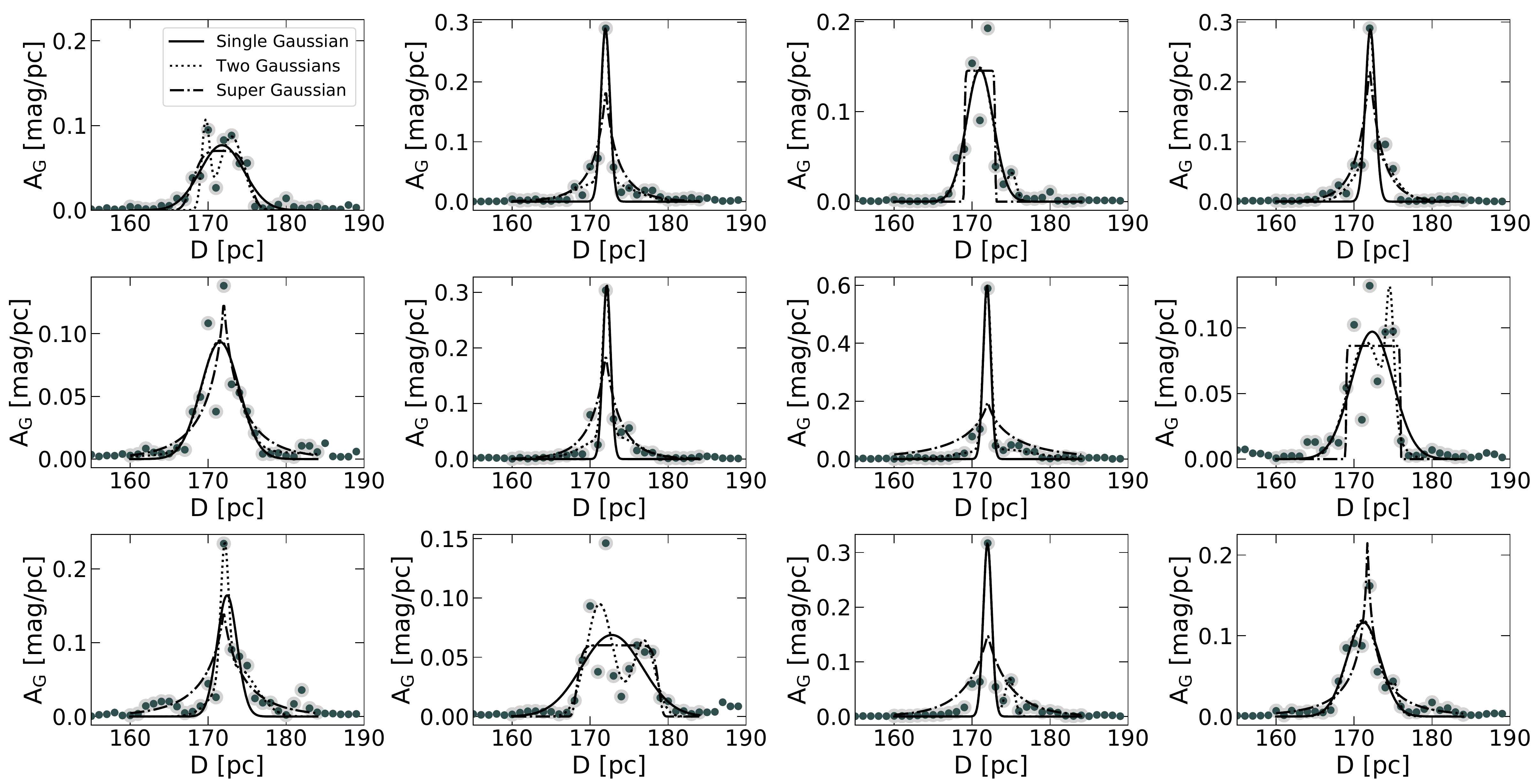}
\caption{The twelve different realizations of differential dust extinction (solid dark points). With the solid, dotted and dash-dotted lines we present our results from fitting a simple Gaussian function, two Gaussian components and a super-Gaussian function, respectively. The data we used to perform our fits are marked as gray-shaded points. While for some realizations, one or more of these parametric models, can be an adequately fit, none of the models is an accurate representation of the data for more that $\sim$50\% of the realizations.
\label{FailedFits}}
\end{figure*}

In Fig.~\ref{FailedFits} we show the results of our efforts to fit three different parametric models to the differential dust-extinction data for a selected sightline with RA, DEC=[186.7083, -71.3484]. The dust-extinction data for each of the twelve different realizations of this sightline are shown as the dark solid points whereas the data points we used to perform our fits are marked as gray-shaded circles. 

First we attempted to fit a single Gaussian component (solid black line in Fig.~\ref{FailedFits}). In at least five different realizations where the fitting of the Gaussian function correctly retrieved the ``peak" in the data its wings miss data points which are closer, but not part, of the background. As a result, the value of the Gaussian function becomes zero before the data. Consequently, even if we considered the depth of Musca to be three time the $\sigma$ of these Gaussian fits, we would still be underestimating the LOS depth of Musca by a few pc. 

This statement also holds true for some realizations where the fit is not driven by the ``peak" in the data (see for instance the bottom left panel of Fig.~\ref{FailedFits}). To make matters worse, the situation is not remedied by considering two Gaussian components instead of one (dotted lines in Fig.~\ref{FailedFits}), or if we fit a super-Gaussian function (dash-dotted lines in Fig.~\ref{FailedFits}). We thus conclude that this approach cannot yield reliable results and an approach where the results do not depend on fitting some parametric model (as described in \S~\ref{dustmaps}) is more reliable.

\end{document}